\documentclass[review]{elsarticle}
\usepackage[utf8]{inputenc}

\usepackage{lineno}
\modulolinenumbers[5]

\journal{Journal of \LaTeX\ Templates}
\usepackage[colorinlistoftodos]{todonotes}
\usepackage{graphicx}
\usepackage{float}
\usepackage{soul}
\usepackage{array}
\usepackage{tabularx}
\usepackage{changes}
\graphicspath{ {./figures/} }
\usepackage{url}

\usepackage{breakurl}
\usepackage[breaklinks]{hyperref}

\begin{document}

\begin{frontmatter}
\title{Observations of peak electric load growth in ERCOT with the rise of electrified heating and its implications for future resource planning}

\author[add1]{Matthew J. Skiles}
\author[add1]{Joshua D. Rhodes, PhD}
\author[add1]{Michael E. Webber, PhD}

\address[add1]{Walker Department of Mechanical Engineering, The University of Texas at Austin, 204 E Dean Keeton St, Austin, TX 78712, USA}

\begin{abstract}
This analysis quantitatively compares the evolution in summer and winter peak demands in the Electric Reliability Council of Texas (ERCOT) service area from 1997 through 2021. Weather data for the days in which peak demand occurred were also compiled to investigate the relationship between peak heating and cooling loads and ambient temperature. This relationship was then applied along with population projections and a climate scenario with medium to high radiative forcing to create winter and summer peak demand growth scenarios for 2025 through 2050. This analysis informs resource planners about how ERCOT peak demand might change in the future and provides new insight into how electric load growth and non-flexible electrified heating demand could have contributed to the February 2021 ERCOT blackouts. We found that historically, summer peak demand growth has been generally stable and approximately linear with time. The stable summer peak load is likely a consequence of fairly constant temperatures observed on summer peak demand days. Conversely, the winter peak demand growth has been less consistent, varying much more around the broader trend. This phenomenon is likely a consequence of high residential electrical heating load on winter peak demand days, which saw temperatures that varied widely from the mean value. Future peak winter and summer electricity demand scenarios indicated that while average temperatures on winter peak demand days will remain fairly constant, they will be more erratic than temperatures on summer peak demand days. As a result, winter peak demand will remain more erratic and will sporadically surpass summer peak demand between 2025 and 2050. Thus, resource planners in ERCOT should place less certainty on winter peak demand projections and an increased level of winter preparedness on both the supply and demand sectors appears warranted. 
\end{abstract}
\end{frontmatter}

\section{Introduction}

Heating is the largest global energy end-use (about 50\% of energy demand) ahead of transport (29\%) and electricity (21\%). Forty-six percent of this heat energy is consumed within buildings for space and water heating and to a lesser extent, cooking \cite{IEA2019}. Natural gas is used to heat 60\% of U.S. households in cold and very cold climates and 47\% of U.S. households overall \cite{lawrence2017}. This end-use consumption of fossil fuels releases significant greenhouse gases from leaks and combustion-related emissions. The electrification of heating is becoming a standard component of decarbonization efforts \cite{sheikh2019,gaur2021}, and bans on fossil fuel based space heating equipment have already been administered in numerous locations \cite{gruenwald2020,derrick2020,DiChristopher2022}. Additionally, legislation such as the Inflation Reduction Act provides tax incentives for electric heat pump installation and might cause rapid adoption of this technology over the next decade. As a result, electrical heating equipment as a portion of global heating technology sales for residential and service buildings has been steadily increasing for years, a trend that is expected to continue \cite{IEA2021}. These factors have led U.S. regional electric grid operators to begin to anticipate a large expansion of space heating demand from the residential and commercial sectors and some to predict a potential switch from a traditional summer peak to a winter peak \cite{MISO2021,NYISO2020}. Questions remain about how this new source of electricity demand will affect electric grid operations. In particular, how will grid resiliency be impacted by the electrification of space heating? How will electric load for residential space heating, which already drives winter peak power demand \cite{ERCOT2021}, and is sensitive to severe weather events, be affected by climate change? Figure 1 demonstrates this phenomenon in one regional electric grid, the Electric Reliability Council of Texas (ERCOT).

\begin{figure}[H]
\includegraphics[scale=0.75]{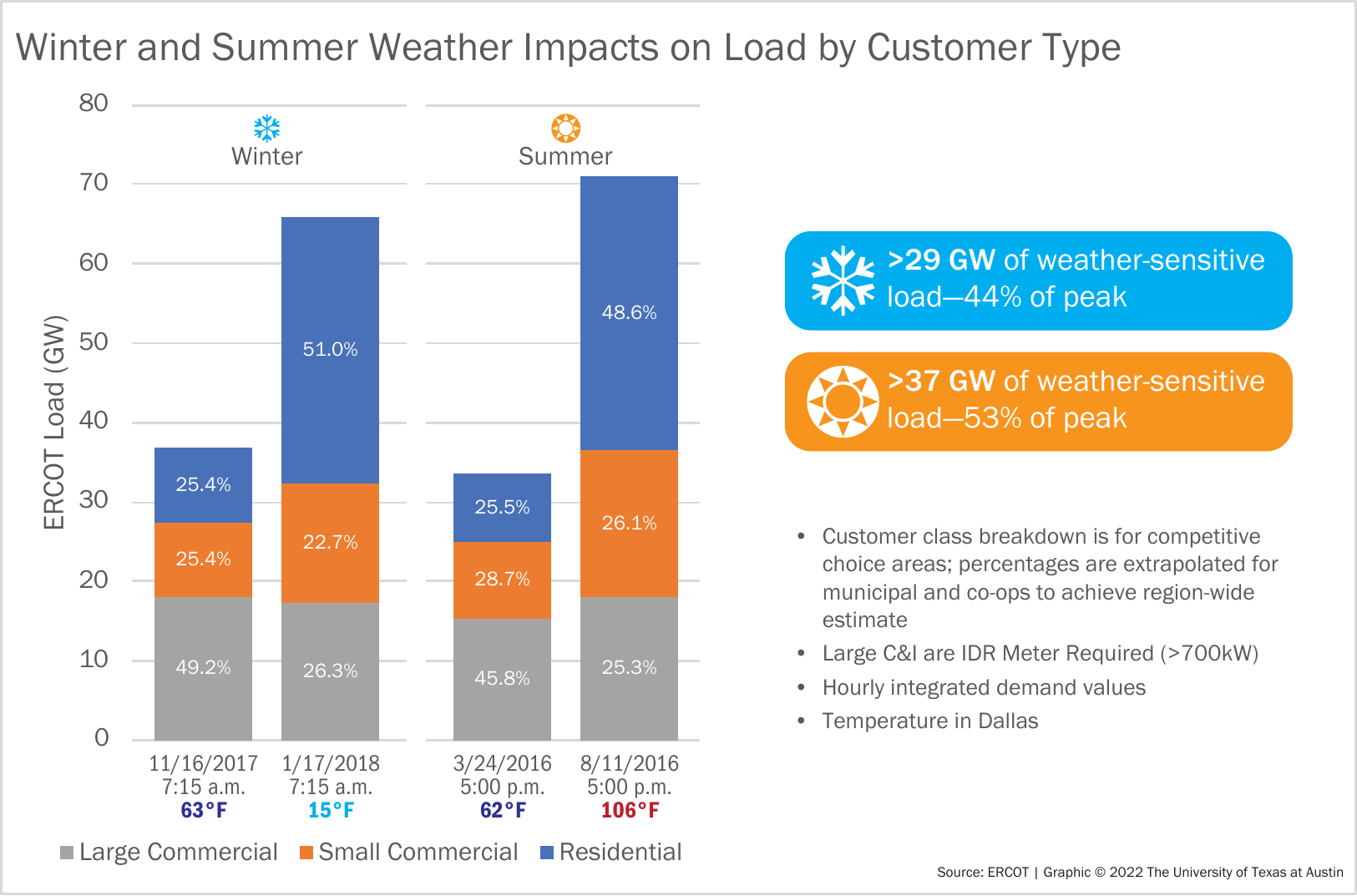}
\caption{The residential sector in ERCOT (the Electric Reliability Council of Texas) is responsible for approximately half of peak demand in the winter and summer \cite{ERCOT2021}.}
\centering
\end{figure}

The ERCOT regional power grid is an ideal test bed for answering questions surrounding the electrification of heating. The ERCOT grid is comprised of a generation mix with high fractions of wind and solar, which much of the U.S. system may soon replicate \cite{reeve2022}. Additionally, a large portion of Texas is in a semi-arid temperate climate and has population centers in hotter and more humid parts of the state. Consequently, electrical heating equipment, which has historically been designed for more moderate temperatures, is prevalent. The percentage of Texas household heating that is met by electricity is increasing over time and is the 7th highest among U.S. states \cite{PUMS2019}. Thus, the demand-side of ERCOT also resembles an evolving, future decarbonized grid. Figure 2 demonstrates the high penetration of electric home heating in Texas relative to the rest of the contiguous United States.

\begin{figure}[H]
\includegraphics[scale=0.3]{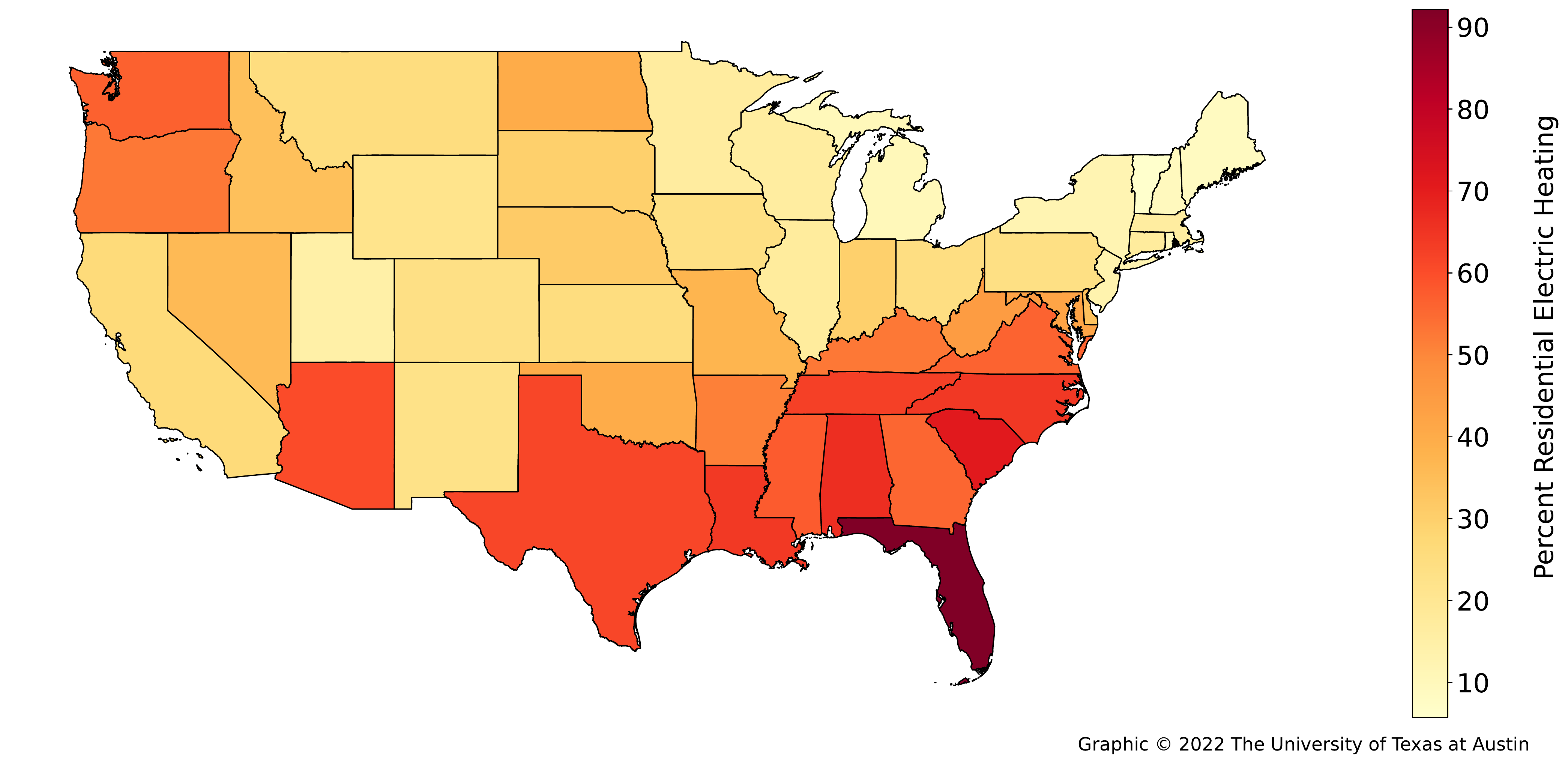}
\caption{As of 2019, Texas had one of the highest rates of electric heating in the residential sector in the United States \cite{PUMS2019}. Generally speaking, electric heating is more prevalent in warmer climates (e.g. the southern part of the United States) or where there is abundant hydroelectric power (e.g. the Pacific Northwest).}
\centering
\end{figure}

As a result of the warm climate, ERCOT resource planners have prioritized managing the natural gas system and electricity grid such that they can meet the demand from large amounts of electrically-operated air-conditioners on hot summer afternoons. However, the events of the winter storm in February 2021, which caused hundreds of deaths and blackouts and boil water notices for millions of people, should serve as a reminder that Texas is not immune to severe winter weather that strains infrastructure reliability \cite{busby2021,glazer2021,king2021,FERC2021}. These winter grid resiliency challenges could yield important lessons for future grid operators in increasingly electrified economies.

The winter event of February 2021 left millions of Texans in ERCOT without power for multiple days during some of the coldest and most widespread winter weather seen in the state in decades. In fact, February 2021 was the first time in recorded history that all 254 counties of Texas were under a winter storm watch at the same time. \cite{villarreal2021}. This event presents the opportunity for a unique case study on a highly electrified region subjected to severe winter weather.

Having lost roughly half of all power generation capacity due to freezing equipment, fuel shortages, and other issues, the grid was minutes away from conditions that could have triggered a total system-wide blackout that then could have required weeks or months for full recovery \cite{king2021}.

Because Texas is the largest energy producing and consuming state in the United States \cite{EIA2022}, it was a surprise to many that the state could run out of energy. As such, there has been continued public outrage and demands for change to prevent a similar disaster. Almost every relevant government agency including the Governor’s office, the State Legislature, Railroad Commission (which is the oil and gas regulator in Texas), the Public Utility Commission of Texas, and ERCOT have faced intense scrutiny or had its top leadership replaced.  Because leaders are under pressure to implement new regulations or market reforms to prevent a future disaster, there is a need to inform the policymaking process in Texas and other regions that may face similar challenges with data about key underlying trends. 

While many reports have assessed the underlying acute causes of the February event itself and its meteorological underpinnings \cite{busby2021,king2021,FERC2021,clack2021,doss-gollin2021}, to the authors’ knowledge, none have rigorously examined how seasonal peak demand has changed over time nor investigated how non-flexible heating electrification could have contributed to the disaster. This analysis seeks to fill that knowledge gap as well as present scenarios for future seasonal ERCOT peak demand growth based on population projections and a future climate scenario. In addition to providing insight into ERCOT grid performance, this analysis presents a relevant case study for a highly electrified economy subject to severe winter weather and thus could have useful insights for grid planners in other regions.

\section{Methods}
The methodology for this analysis is simple in principle yet allows for some important observations. Peak demand data for the past 25 years in ERCOT were compiled for the winter (December through February) and summer (June through August) seasons \cite{ERCOT,ERCOT2009,FERCandNERC2011,king2021,ERCOTb}\footnote{2001 summer peak demand sourced from \cite{ERCOTb} as "July Load at ERCOT Coincident Peak kW".}. A linear regression was used to assess their growth over the 25-year period and the effect of ambient temperatures on these events was investigated. 

\paragraph{Weather influence on peak demand} Degree day data were calculated to assess the sensitivity of peak demand to outdoor temperatures. A degree day compares the mean outdoor temperature for a location to a base temperature as a measure of cooling or heating load \cite{EIA2021}. Cooling degree days (CDD) and heating degree days (HDD) are calculated using ambient temperature readings collected at a particular location throughout the day. The time in days between two temperature readings is multiplied by the number of degrees by which the ambient temperature was above or below the base temperature over the period to get the degree days \cite{degreedaysa}. The further the ambient temperature is above or below the base temperature, the higher the CDD or HDD respectively, and thus the higher the cooling or heating load.
\\
\begin{equation}
CDD = \sum time\; between\; readings\; (days) \times (ambient\; temp\; -\; base\; temp)
\end{equation}
\\
\begin{equation}
HDD = \sum time\; between\; readings\; (days) \times (base\; temp\; -\; ambient\; temp)
\end{equation}
\\

In this study, we calculated the number of degree days during each peak demand day. The base temperature for calculating both CDD and the HDD was set to 18.5° C (65.3° F) and an ERCOT-wide degree day (DD) value was calculated by taking DD values from the largest city in each ERCOT weather zone \cite{degreedaysb,wunderground} and weighting each DD value by the population in the weather zone\footnote{2021 population projected based on 2019-2020 percent population growth.} \cite{uscensus} as described by Equations 1-3 and Figure 4. 
\\
\begin{equation}
ERCOT\; DD=\sum DD\; from\; largest\; city\; in\; each\; zone\times \frac{zone\; population}{ERCOT\; population}
\end{equation}
\\

\begin{figure}[H]
\includegraphics[scale=1]{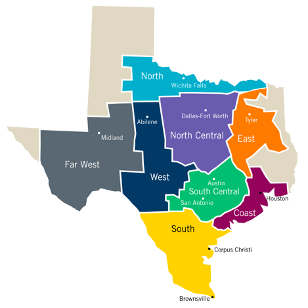}
\caption{ERCOT uses eight weather zones for planning purposes \cite{ERCOTc}.}
\centering
\end{figure}

\begin{figure}[H]
\includegraphics[scale=0.4]{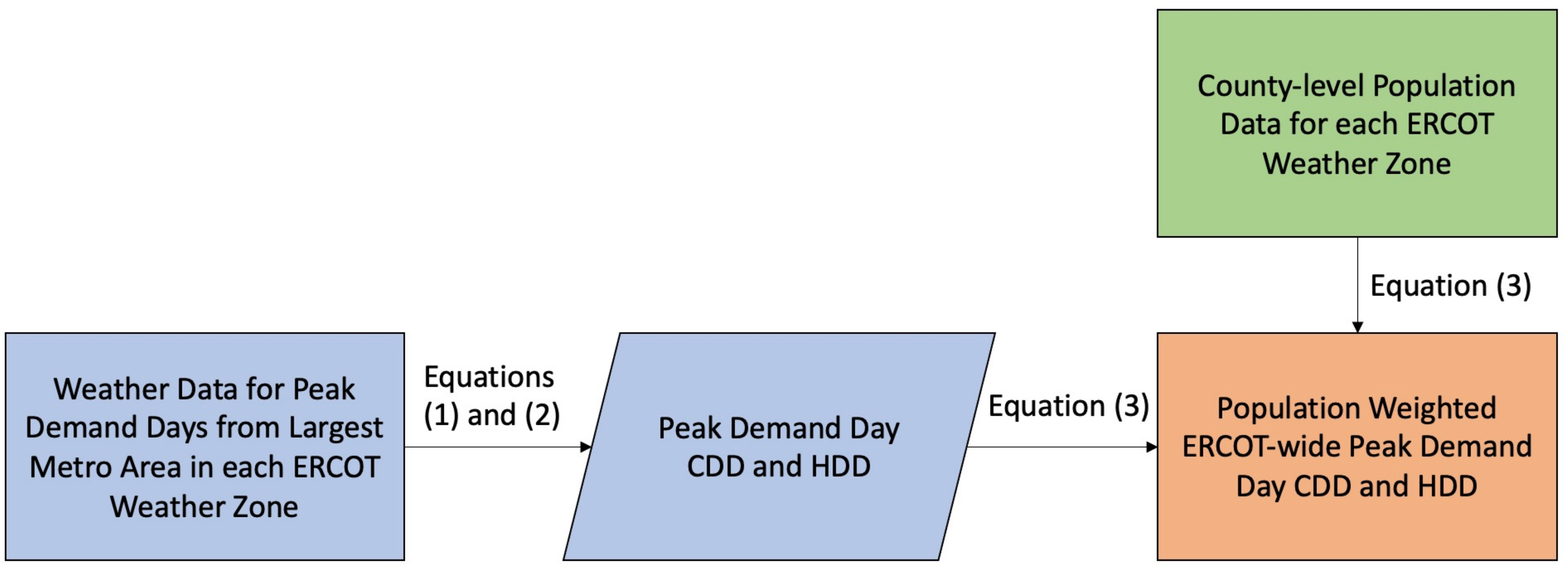}
\caption{This flow diagram shows the methodology used to calculate population-weighted CDD and HDD in ERCOT.}
\centering
\end{figure}

\paragraph{Future CDD and HDD scenarios}Future CDD and HDD were calculated using daily average near-surface temperature as processed from downscaled global climate models produced by the NASA Center for Climate Simulation (NCCS) \cite{NASAEarthExchange}. NCCS describes the dataset as ``comprised of global downscaled climate scenarios derived from the General Circulation Model (GCM) runs conducted under the Coupled Model Intercomparison Project Phase 6 (CMIP6) and across two of the four “Tier 1” greenhouse gas emissions scenarios known as Shared Socioeconomic Pathways (SSPs)" \cite{NASAEarthExchange}. This dataset was developed for the Sixth Assessment Report produced by the Intergovernmental Panel on Climate Change. The dataset chosen for this analysis was the SSP3-7.0 scenario, which represents the medium to high end of the range of future radiative forcing pathways \cite{abram2019}.

\section{Results and discussion}

Figure 5 shows how the winter\footnote{Winter peaks in 2011 and 2021 were estimated because firm load shedding prevented the full load from being served. The peaks for those years were taken from ERCOT estimates of what load would have been absent load shed. Winter peak in 2001 was taken from an ERCOT report as 2001 load data were not available.} and summer\footnote{Summer peak in 2001 was taken from ERCOT Coincident Peak Calculations as 2001 load data were not available.} peaks have grown relative to each other along with a linear regression for each season. General observations of the peak demand linear regression results yield two main conclusions; 1) ERCOT’s winter peak is growing about 15\% faster than its summer peak, based on the slopes of the linear regressions and 2) the winter peak is more erratic than the summer peak, based on the lower R-squared values of the same regression. The winter peak is, on average, about 3.5 GW off (above or below the mean of the absolute values of the regression model errors) the linear model estimation, while the summer peak is only about 1 GW off, on average. Figure 6 shows how each year’s summer and winter peaks differ from the linear model’s prediction.

\begin{figure}[H]
\includegraphics[scale=0.75]{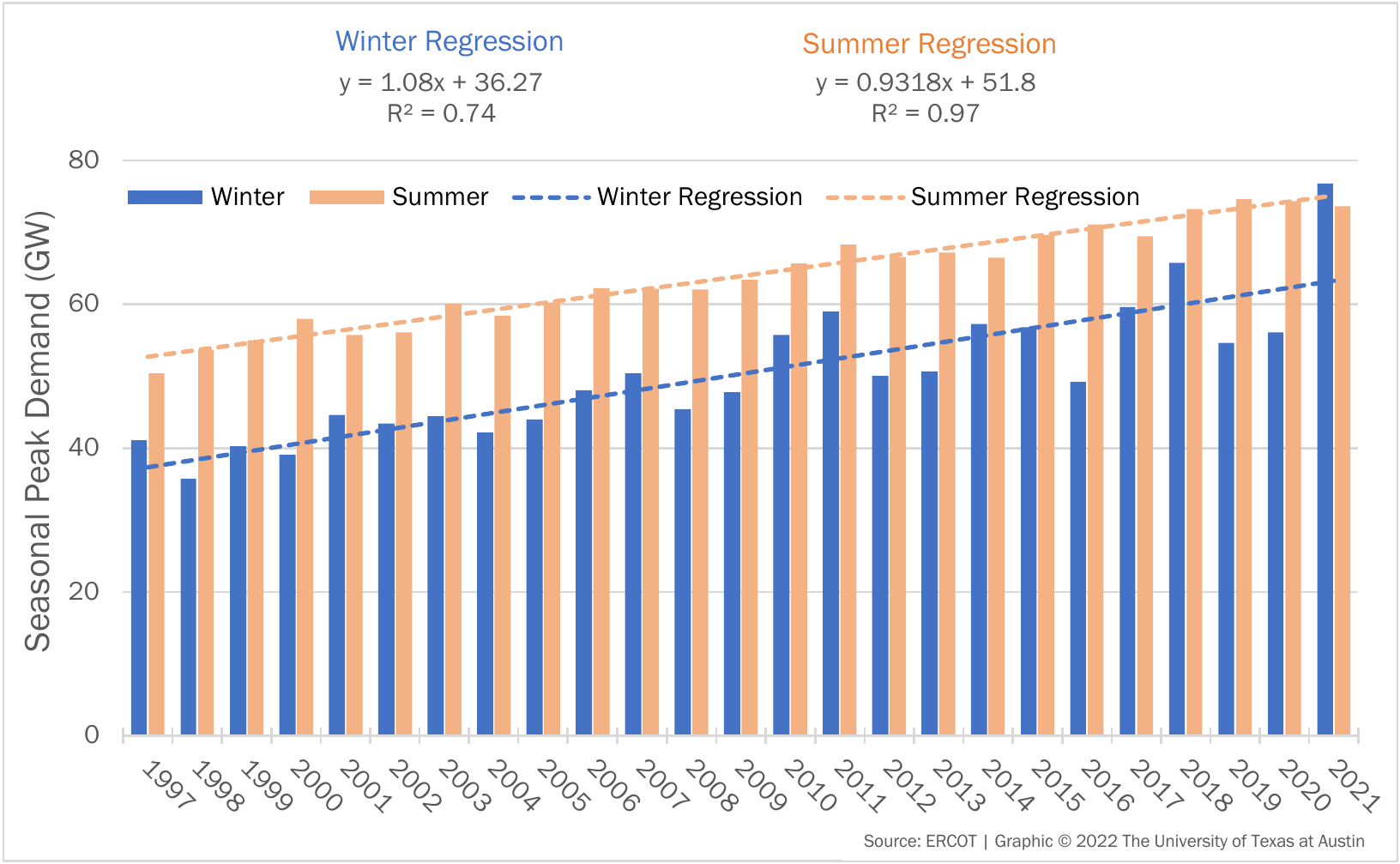}
\caption{Winter and summer peak electrical demand in ERCOT grew from 1997 to 2021. The bars are actual peak demands (or estimated peak demands if load shed happened) and the dotted lines are the linear fit estimations of peak demand for each season and each year.}
\centering
\end{figure}

\begin{figure}[H]
\includegraphics[scale=0.75]{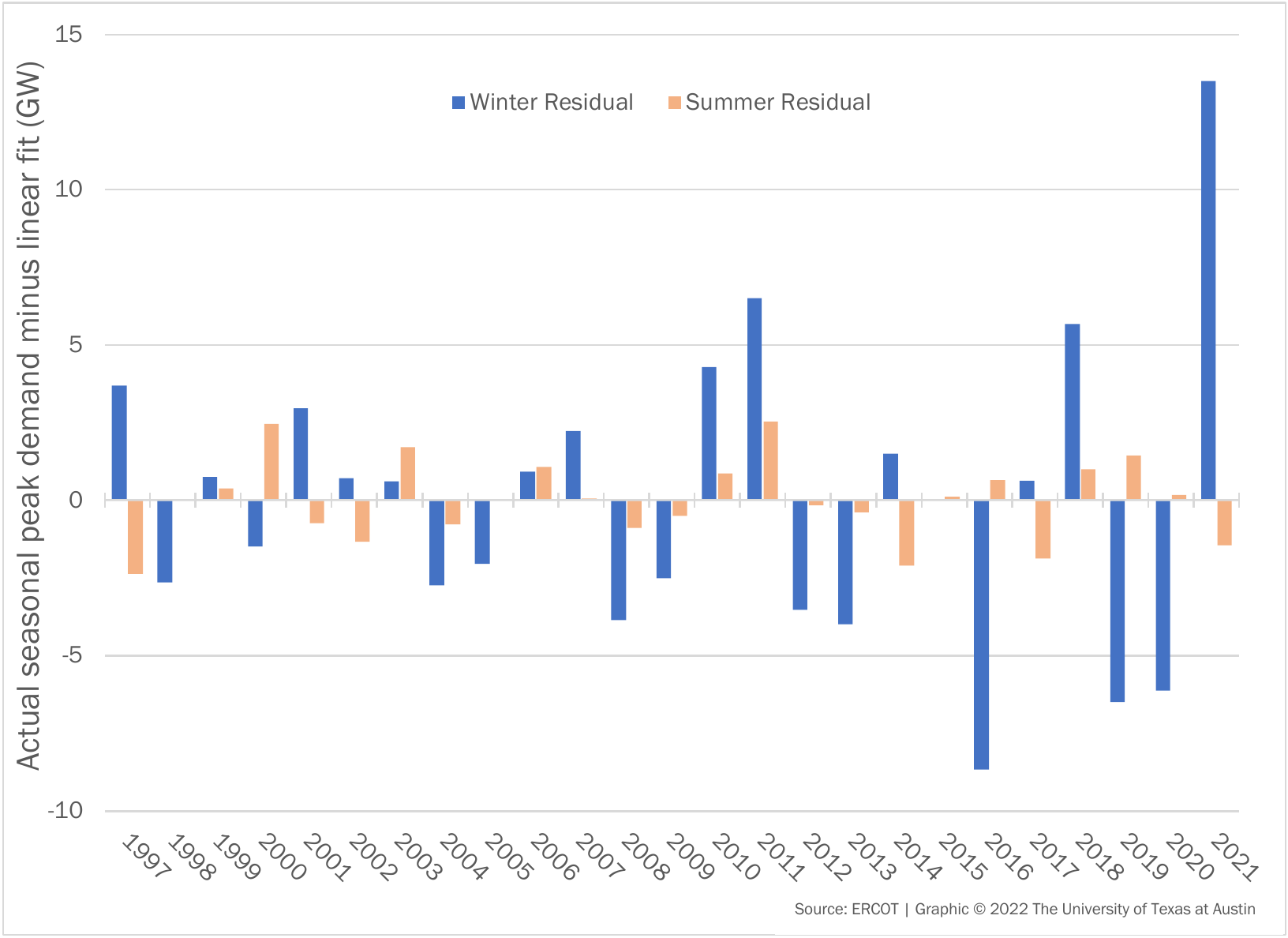}
\caption{The difference between the estimated seasonal peak and the actual peak for each year shows that winter peak demand is much more variable (and therefore harder to predict) than summer peak demand.}
\centering
\end{figure}

The linear fit in Figure 5 shows that the summer peak demand growth is more “stable” (or consistent) than the winter peak. A visual inspection of the fit (dotted lines) versus the actual\footnote{Estimated peak demands in case of any firm load shed.} peak demands (bars) shows that the summer peak growth is more linear than the winter model. This conclusion is also indicated by the R-squared values of the respective models (higher R-squared values correspond to higher linearity): R-squared value of 0.74 in the winter vs 0.97 in the summer.

The lower R-squared value of the winter model indicates that the growth in winter peak is less predictable. The summer peak is presumably a combination of consistent population and economic growth trends offset somewhat by efficiency improvements to space conditioning appliances. The winter peak has similar factors with two confounding weather factors from climate change: slowly increasing average prevailing temperatures (which reduces average total seasonal energy for heating) and ongoing risk for intense winter storms.

Since the general formation of ERCOT, the annual winter peak has never exceeded the summer peak. If the current rate of change shown in Figure 5 continues, it would take over 100 years from the start of the analysis (1997) for the average winter peak to systematically surpass the average summer peak. However, if the ERCOT grid had been able to deliver as much power as was estimated to have been demanded in the winter storm of February 2021 \cite{king2021} the winter peak would have surpassed the summer peak for the first time since the Texas grid operator began recording data for its modern grid footprint. Furthermore, if electrified heating deployment rates accelerate, then the summer peak demand will grow less quickly (because of commensurate upgrades to air conditioning efficiency) and the winter peak demand will grow more quickly, in which case it is reasonable to anticipate that the winter peak will regularly exceed the summer peak much sooner than a century from 1997.

Because of the higher variability, peak demand projections for ERCOT, which are generally based on a “normal weather year” and are generally linear \cite{ERCOT2021b}, should include more uncertainty on the forecasted winter peak than the summer peak.

While the summer cooling season ramps up and down over several months, the winter heating season is much more erratic. For homes with electric heating, the increase in demand for home heating when outside temperatures drop to levels seen in February 2021 is larger than the demand is for home cooling when a heat wave pushes temperatures to summer peak levels \cite{NRG2021}.

While the residential sector is generally responsible for most of the large swings in demand, residential demand response in ERCOT is small relative to commercial and industrial demand response \cite{du2019}. As such, residential demand response programs that seek to reduce peak demand are mostly an untapped potential solution for grid reliability. 

\paragraph{Peak demand sensitivity to weather}Degree day data indicate that, over the past 25 years, temperatures were more erratic on winter peak demand days than on summer peak demand days. CDD on summer peak demand days were fairly constant near the mean CDD value. However, HDD for winter peak demand days varied widely from the mean value (Figure 7). The unpredictable nature of the temperatures on winter peak demand days and more consistent temperatures on summer peak demand days is mirrored by the heating and cooling loads, respectively.

\newcolumntype{P}[1]{>{\raggedright\arraybackslash}p{#1}}
\newcolumntype{M}[1]{>{\centering\arraybackslash}m{#1}}
\begin{table}
  \centering
  \begin{tabular}{|M{2.3cm}|M{2.3cm}|M{1.7cm}|M{2.7cm}|M{2.7cm}|} 
    \hline
    Weather Zone & City & Weather Station ID & Average Summer Peak Demand Day CDD & Average Winter Peak Demand Day HDD \\ \hline\hline
    Coast & Houston & KIAH & 12.5 & 16.2 \\ \hline
    East & Tyler & KTYR & 13.0 & 19.8 \\ \hline
    Far West & Midland & KMAF & 11.9 & 20.9 \\ \hline
    North & Wichita Falls & KSPS & 13.6 & 22.6 \\ \hline
    North Central & DFW & KDFW & 14.8 & 20.4 \\ \hline
    South & Corpus Christi & KCRP & 11.6 & 13.4 \\ \hline
    South Central & Austin & KAUS & 12.6 & 17.6 \\ \hline
    West & Abilene & KABI & 12.7 & 21.6 \\ \hline
  \end{tabular}
  \caption{Summary of weather stations used in calculation of state-wide population-weighted CDD and HDD values and average CDD and HDD values.}\label{tab1}
\end{table}

\begin{figure}[H]
\includegraphics[scale=0.75]{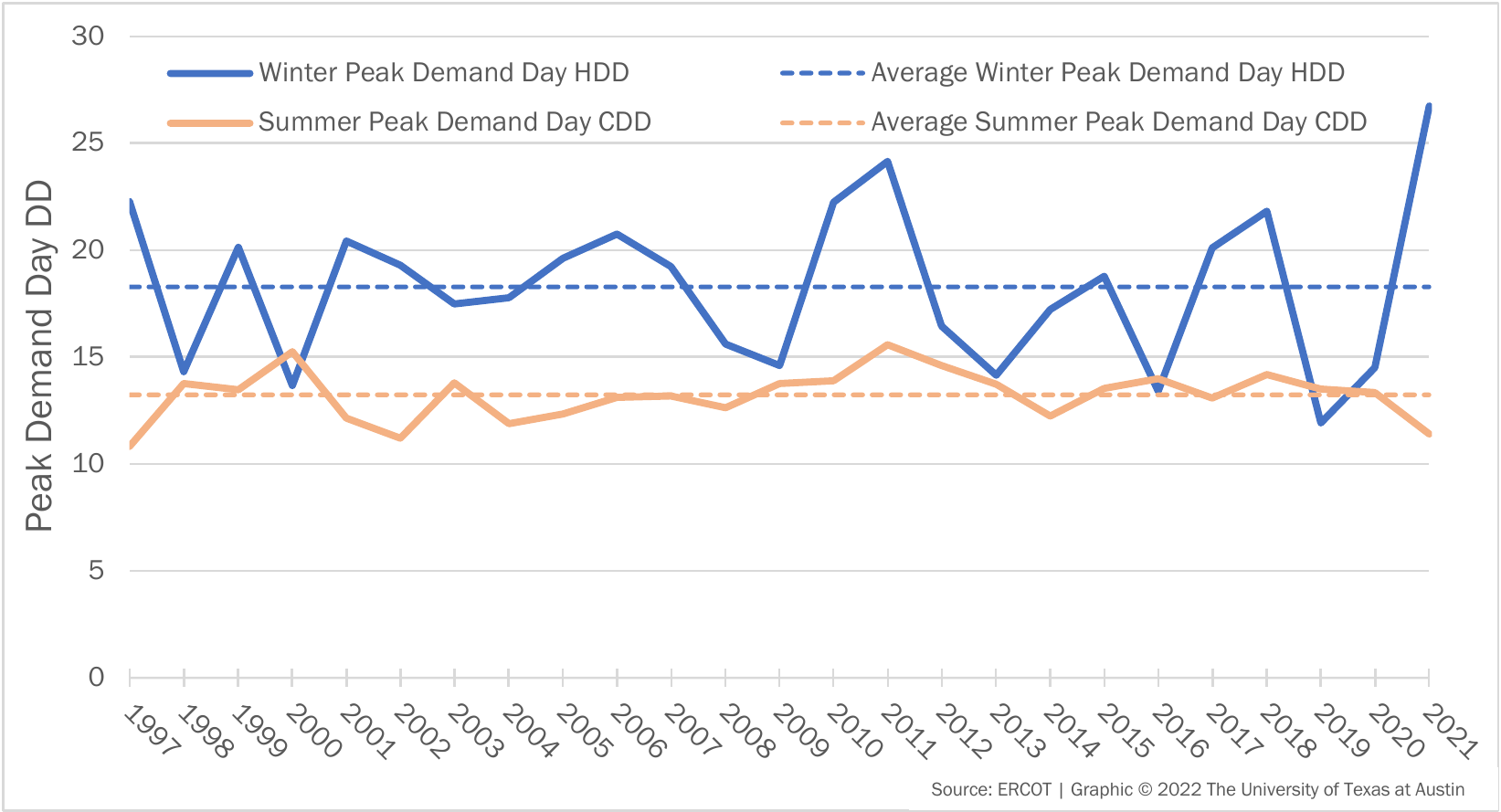}
\caption{HDD on winter peak demand days are more variable than CDD on summer peak demand days. ERCOT-wide DD values are calculated from DD values for the largest population center in each ERCOT weather zone weighted by the population of each weather zone.}
\centering
\end{figure}

These weather trends that drive heating and cooling loads impact power grid operations. Overall, about 55\% of residential heating equipment in the West South Central census region of the US, which includes Texas, is electric. Of those electric heaters, about 85\% are electric resistance and the remaining 15\% are heat pumps \cite{EIA2018}. Electrical resistance heating has very high power draws as compared to other forms of heating, such as heat pumps. Heat pumps can provide heating using much less electricity most of the time. However, many of the heat pumps installed in the region are not cold-weather heat pumps, rather they are usually designed for mild climates, and thus are only able to operate down to a specified low outdoor temperature before switching to backup or auxiliary heating modes that rely heavily on electric resistance heating or natural gas \cite{white2019,white2021}. When this switch happens, it can create a large jump in the electricity demand of each individual heating system that in aggregate can create a large upward disruption in grid demand as a cold front moves through. An analysis of per capita winter peak demand versus HDD exhibits this effect: per capita winter peak demand appears to have a polynomial relationship with degree days, becoming increasingly high for higher HDD (Figure 8). Additionally, data points from more recent years with higher penetration of electric heating tend to reside above the regression line, indicating that more electricity is consumed per capita when more electrical heating equipment is connected to the grid.

\begin{figure}[H]
\includegraphics[scale=0.5]{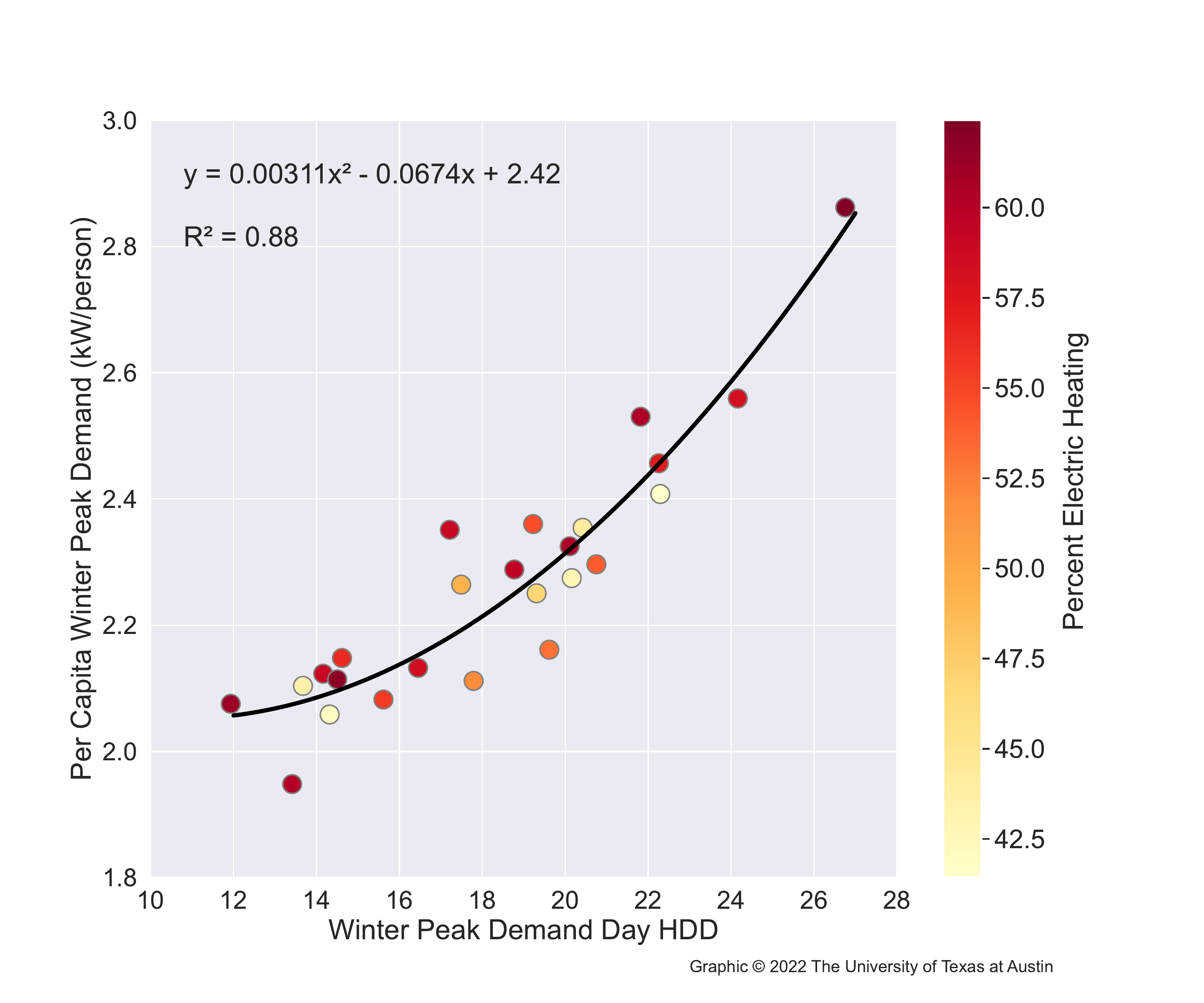}
\caption{The relationship between 1997-2021 per capita winter peak demand and winter peak demand day HDD appears to be polynomial. Percent heating electrification data for this figure was sourced from the American Community Survey \cite{PUMS2019} and the Residential Energy Consumption Survey \cite{EIA2018} and interpolated or extrapolated for years in which data was not available.}
\centering
\end{figure}

Normalizing the peak demand data by population and weather demonstrates that the summer per-capita, per cooling degree day peak demand is decreasing, indicating that the electrical efficiency of meeting the summer peak demand is increasing (Figure 9).

However, the winter per-capita, per heating degree day peak demand is increasing, indicating that over time progressively more electricity is being consumed to meet the same per capita heating load. However, unlike the summer peak demand which has been driven by electrically-powered air-conditioners for the entire period of our dataset, the makeup of heating equipment has changed over time. For example, in 1997, only about 40\% of homes in Texas used electrical heating, but that number has increased over time, meaning that the percentage of homes heating with other fuels, such as natural gas, has declined \cite{EIA2018,PUMS2019}. Analyzing the flows of natural gas is beyond the scope of this analysis, but the positive slope, in Figure 9, of the winter per-capita, per heating degree day peak demand linear fit does indicate (although weakly) that electricity use in the winter is generally increasing even when population and weather are considered. This is likely capturing the dual impacts of heat pumps replacing electrically-operated AC units and gas furnaces. By replacing older AC units with a new heat pump, the electrical efficiency of cooling improves. By replacing gas furnaces with heat pumps, heating load is shifted from natural gas infrastructure to the power grid.

\begin{figure}[H]
\includegraphics[scale=0.77]{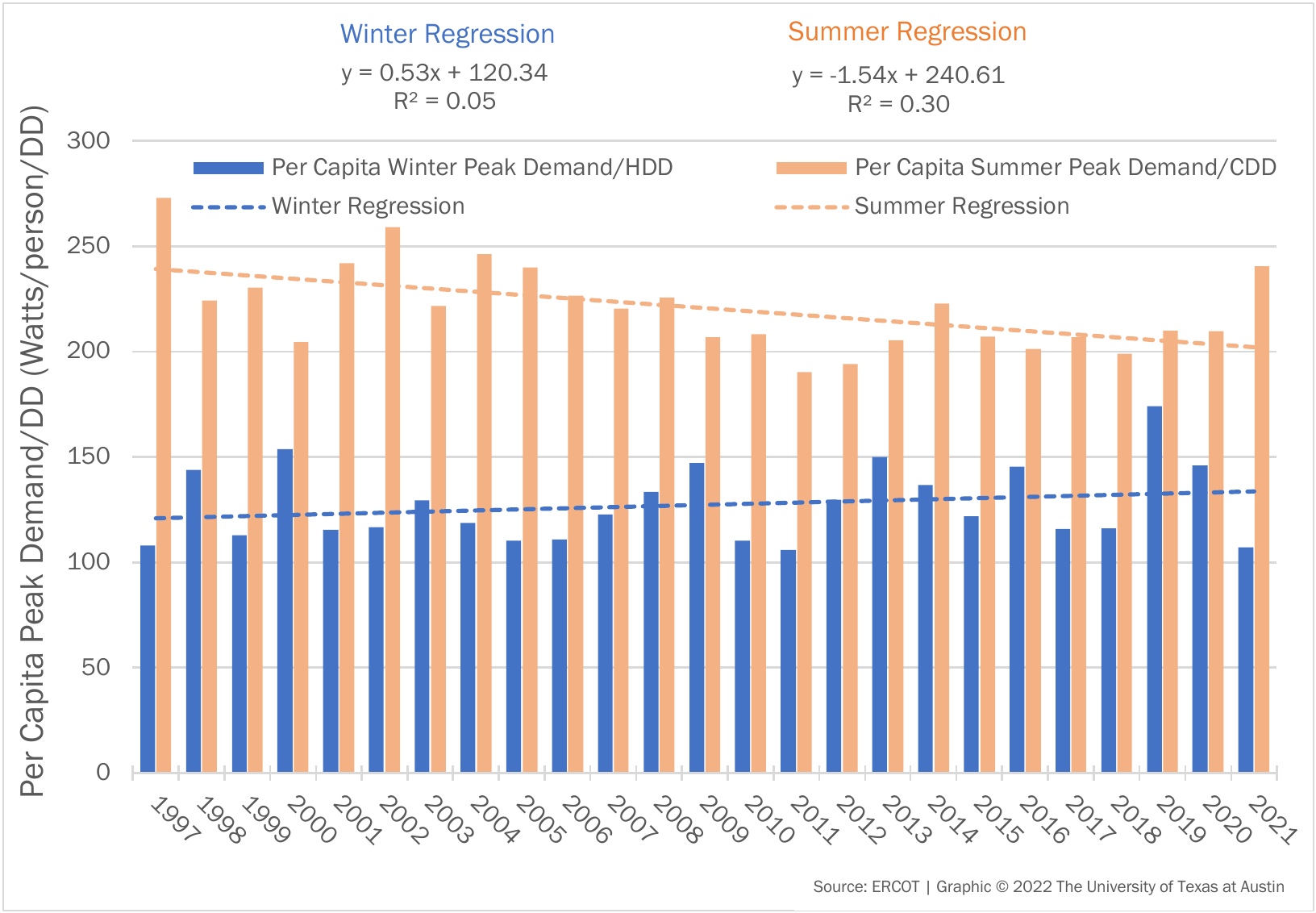}
\caption{Per capita peak demand per DD is decreasing over time for summer peak demand days and increasing over time for winter peak demand days.}
\centering
\end{figure}

\paragraph{Future peak demand scenarios} Scenarios of winter and summer peak demand were constructed by extrapolating the per-capita, per degree day peak demand linear fits out for 2025 through 2050 and utilizing population projections \cite{2022StatePlan} and DD for a future climate scenario (Figure 10). In this peak demand scenario for 2025 through 2050, winter peak demand day HDD and winter peak demand continue to be more erratic than summer peak demand day CDD and summer peak demand. Additionally, while average winter peak demand day HDD remains about the same, average summer peak demand day CDD rises by $\sim$15\% as temperatures rise through mid-century. Summer peak demand through 2050 is more erratic than it has been historically, but is still more stable than winter peak demand through 2050 and is generally increasing at a slower rate than winter peak demand due to changing heating and cooling efficiencies that arise due to the installation of electrical heating equipment.

\begin{figure}[H]
\includegraphics[scale=0.77]{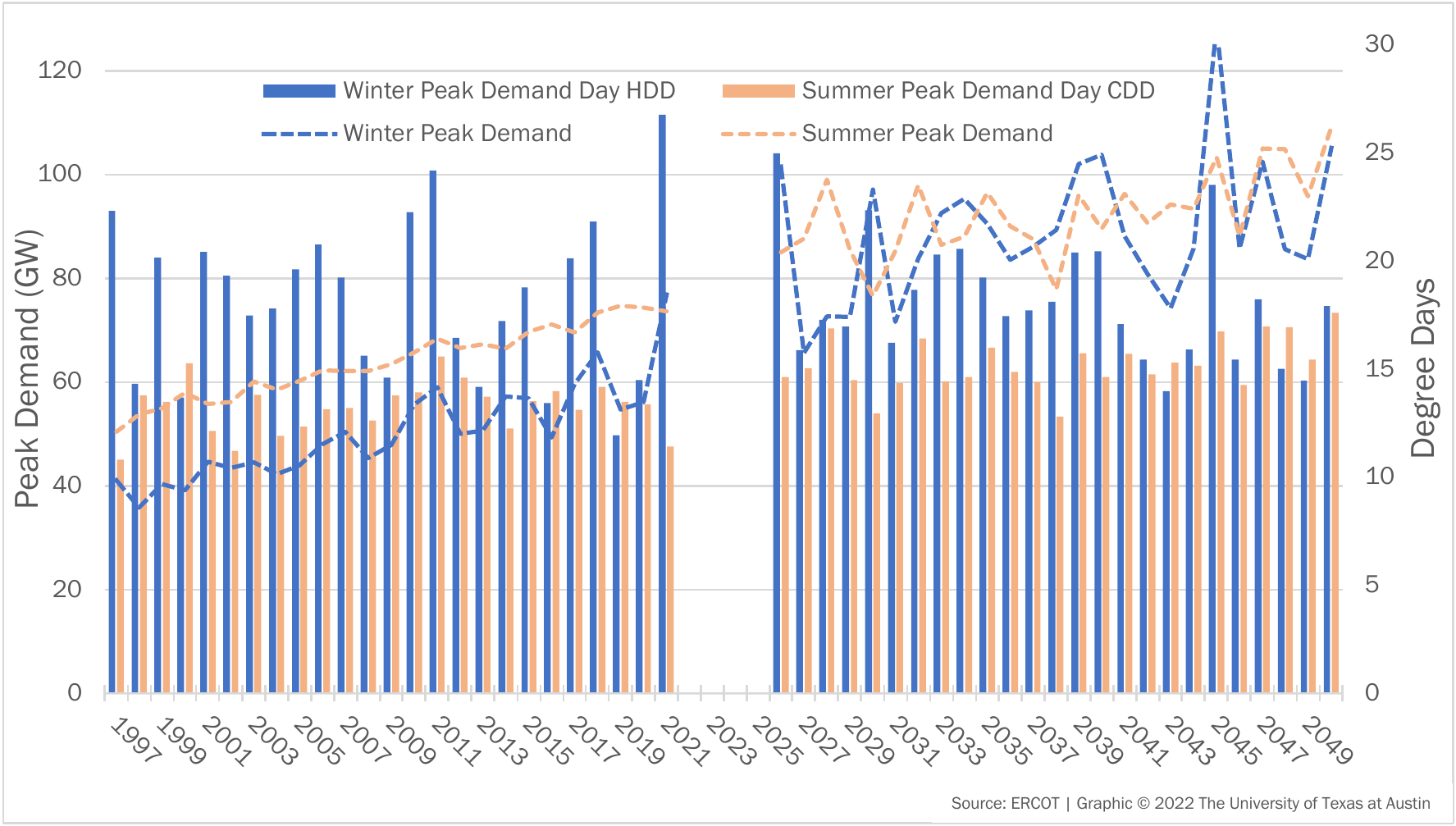}
\caption{Winter and summer peak demand scenarios through 2050 based on extrapolated historical per capita peak demand per DD with population projections and DD calculated for a future climate scenario demonstrate that winter peak demand will likely continue to be more erratic than summer peak demand, and winter peak demand will begin to regularly surpass summer peak demand before 2050.}
\centering
\end{figure}

\paragraph{Suggestions for system and resource planning} As Texas and other regions electrify heating, more attention will need to be given to the impacts of extreme periods of cold weather on winter peak demand. Increased efficiency and cold-weather standards for heat pumps could allow the continued electrification of heating to avoid making as big of an impact on winter peak demand by delaying the individual heater’s switch into auxiliary (electric resistance) mode, thereby reducing how often demand grows polynomially.

Increased residential building envelope efficiency could also reduce the amount of heating needed by the average home which would reduce the amount of individual coincident heating systems operating and thus lower overall heating demand from the residential sector. Most of Texas is located in “hot” climate zones and thus more attention has been paid to constructing homes to withstand high summer temperatures. Homes in Texas are generally designed to have adequate insulation levels and heating systems to withstand -4C (25F) [30] and are thus not prepared to cope with the much colder temperatures from events such as the winter storm in February 2021.

On the demand side, Texas can avert future load shed by improving its mediocre energy efficiency performance as a cheaper alternative to building new thermal power generation \cite{Lee2022}. It's also worth noting that because half of the peak demand in winter and summer is from space heating and cooling, these end-uses represent a significant opportunity for demand response (e.g. dispatching an intentional reduction in peak demand by turning off heating or cooling devices). These factors have led grid operators to posit that water heating and HVAC are good candidates for load control, and could allow buildings to provide grid services as thermal batteries \cite{MISO2021}. If equipment with flexible features is installed, then rotating shut-offs to electric air conditioners and heaters can help preserve grid reliability and can be cheaper and quicker to install than to build additional generating supply. 

On the grid supply side, it should be noted that by the end of 2022 it is possible that 4-5 GW of Texas power generation (or $\sim$5\% of total Texas generation capacity) will come from the distribution grid \cite{lewin2022}. These distributed energy resources (DERs) could provide valuable grid services if properly coordinated. Price signals and/or incentives should be administered to ensure that DERs are optimally placed to provide grid reliability and cost benefits \cite{lewin2022}. Additionally, increased attention to resource adequacy levels in the winter should be a focus for future operations. While beyond the scope of this work, it is possible that the addition of a new winter reliability product for the ERCOT market might be necessary to meet the additional challenges that are associated with the increased uncertainty in predicting the winter peak demand.

This analysis corroborates previous NERC conclusions that grid planners can have reasonable confidence that each summer in ERCOT will be hot, but should place less certainty around how cold each winter will be\added{ \cite{NERC2022}}. These weather trends hold true historically, and are likely to continue for future climate scenarios through 2050. These findings would indicate the ERCOT grid might need more reserve capacity (higher reserve margin) to handle higher levels of uncertainty in peak demand in the wintertime than in the summer, particularly if the electrification of heating continues. 

Currently, generation maintenance outages for generators and transmission assets are restricted between May 15 and September 15 so that most assets are ready to meet the summer peak demand. However more attention might need to be placed on having more capacity online during winter months to cover the higher levels of uncertainty. But squeezing more generation outages into smaller periods of time might then create supply shortages in the shoulder months.

\section{Conclusions}

This analysis assessed the relative change in historical summer vs. winter peaks in ERCOT from 1997 through 2021 and scenarios of peaks from 2025 through 2050. It was shown that while historical summer peak demand growth is relatively stable, the winter peak demand is growing more quickly and more erratically. This trend is likely the result of the electrical efficiency increases for cooling and growth in electricity consumption as a result of the installation of electrical heating equipment. Additionally, it was shown that the historical instability of the winter peak demand growth is a consequence of erratic winter weather and electrical heating that becomes increasingly inefficient at lower temperatures. When these trends in heating and cooling efficiency are extrapolated through 2050 and combined with population projections and a future climate scenario, winter peak demand continues to grow more quickly and erratically than summer peak demand. This results in winter peak demand regularly surpassing summer peak demand before 2050. These results imply that special attention will need to be paid to how ERCOT plans for winter heating seasons, particularly given the trend of heating electrification in the residential sector, which was already the swing consumer of electricity in both the summer and winter seasons. Namely, planners should use higher uncertainty in their estimates for peak winter demand. Increased general efficiency standards for buildings and in particular heat pumps could mitigate some of the demand side winter issues and an increased level of uncertainty placed on the aggregate electricity grid winter peak demand estimates could help drive policies to increase winter reserves. Considering that the electrification of heating is a primary component of decarbonization efforts in all climates, lessons from this case study should be considered in colder regions as they seek to replace fossil fuel heating equipment with electric heating.  

\section{CRediT authorship contribution statement}
\textbf{Matthew J. Skiles:} Data curation; Formal Analysis; Investigation; Methodology; Visualization; Writing --- original draft, Writing --- review \& editing. \textbf{Joshua D. Rhodes:} Conceptualization; Data curation; Formal analysis; Investigation; Methodology; Visualization; Writing --- original draft, Writing --- review \& editing. \textbf{Michael E. Webber:} Supervision; Writing --- review \& editing; Project administration.

\section{Acknowledgements}

We would like to thank Gene Preston, whose question about the relative changes in  ERCOT’s seasonal peak demand prompted this analysis. This work was supported by the Texas State Energy Conservation Office with additional partial support from the U.S. Army Corps of Engineers’ Engineer R\&D Center (ERDC) through a subcontract with Artesion, the Energy Foundation, and Austin Energy. This material is based upon work supported by the National Science Foundation Graduate Research Fellowship Program under Grant No. DGE-1610403. Any opinions, findings, and conclusions or recommendations expressed in this material are those of the author(s) and do not necessarily reflect the views of the National Science Foundation. The sponsors had no direct involvement in this work. The authors declare that they have no known competing financial interests or personal relationships that could have appeared to influence the work reported in this paper. In addition to the research work on topics generally related to energy systems at the University of Texas at Austin, one of the authors (Webber) has an affiliation with Energy Impact Partners (a venture investment firm) and two of the authors (Webber and Rhodes) are partners in IdeaSmiths LLC (an engineering consulting firm). Any opinions, findings, conclusions or recommendations expressed in this material are those of the authors and do not necessarily reflect the views of the sponsors, Energy Impact Partners, or IdeaSmiths LLC. The terms of this arrangement have been reviewed and approved by the University of Texas at Austin in accordance with its policy on objectivity in research.

\biboptions{sort&compress}
\bibliographystyle{elsarticle-num}
\bibliography{main}

\begin{thebibliography}{10}
\expandafter\ifx\csname url\endcsname\relax
  \def\url#1{\texttt{#1}}\fi
\expandafter\ifx\csname urlprefix\endcsname\relax\def\urlprefix{URL }\fi
\expandafter\ifx\csname href\endcsname\relax
  \def\href#1#2{#2} \def\path#1{#1}\fi

\bibitem{IEA2019}
\href{https://www.iea.org/reports/renewables-2019}{Renewables 2019 –
  {Analysis}}.
\newline\urlprefix\url{https://www.iea.org/reports/renewables-2019}

\bibitem{lawrence2017}
C.~Lawrence, C.~Berry,
  \href{https://www.eia.gov/todayinenergy/detail.php?id=30672}{U.{S}.
  households’ heating equipment choices are diverse and vary by climate
  region} (Apr. 2017).
\newline\urlprefix\url{https://www.eia.gov/todayinenergy/detail.php?id=30672}

\bibitem{sheikh2019}
I.~Sheikh, D.~Callaway,
  \href{https://www.proquest.com/docview/2309963397/abstract/8223CA650C334BC3PQ/1}{Decarbonizing
  {Space} and {Water} {Heating} in {Temperate} {Climates}: {The} {Case} for
  {Electrification}}, Atmosphere 10~(8), place: Basel, Switzerland Publisher:
  MDPI AG (Aug. 2019).
\newblock \href {https://doi.org/https://doi.org/10.3390/atmos10080435}
  {\path{doi:https://doi.org/10.3390/atmos10080435}}.
\newline\urlprefix\url{https://www.proquest.com/docview/2309963397/abstract/8223CA650C334BC3PQ/1}

\bibitem{gaur2021}
A.~S. Gaur, D.~Z. Fitiwi, J.~Curtis,
  \href{https://linkinghub.elsevier.com/retrieve/pii/S221462962030339X}{Heat
  pumps and our low-carbon future: {A} comprehensive review}, Energy Research
  \& Social Science 71 (2021) 101764.
\newblock \href {https://doi.org/10.1016/j.erss.2020.101764}
  {\path{doi:10.1016/j.erss.2020.101764}}.
\newline\urlprefix\url{https://linkinghub.elsevier.com/retrieve/pii/S221462962030339X}

\bibitem{gruenwald2020}
T.~Gruenwald, M.~Lee,
  \href{https://rmi.org/2020-watt-a-year-for-building-electrification/}{2020:
  {Watt} a {Year} for {Building} {Electrification}!} (Dec. 2020).
\newline\urlprefix\url{https://rmi.org/2020-watt-a-year-for-building-electrification/}

\bibitem{derrick2020}
A.~Derrick,
  \href{https://durkan.seattle.gov/2020/12/mayor-durkan-announces-ban-on-fossil-fuels-for-heating-in-new-construction-to-further-electrify-buildings-using-clean-energy/}{Mayor
  {Durkan} {Announces} {Ban} on {Fossil} {Fuels} for {Heating} in {New}
  {Construction} to {Further} {Electrify} {Buildings} {Using} {Clean} {Energy}}
  (Dec. 2020).
\newline\urlprefix\url{https://durkan.seattle.gov/2020/12/mayor-durkan-announces-ban-on-fossil-fuels-for-heating-in-new-construction-to-further-electrify-buildings-using-clean-energy/}

\bibitem{DiChristopher2022}
T.~DiChristopher,
  \href{https://www.spglobal.com/marketintelligence/en/news-insights/latest-news-headlines/gas-ban-monitor-west-coast-pushes-new-boundaries-pro-gas-state-bills-stall-69969602}{Gas
  {Ban} {Monitor}: {West} {Coast} pushes new boundaries; pro-gas state bills
  stall} (Apr. 2022).
\newline\urlprefix\url{https://www.spglobal.com/marketintelligence/en/news-insights/latest-news-headlines/gas-ban-monitor-west-coast-pushes-new-boundaries-pro-gas-state-bills-stall-69969602}

\bibitem{IEA2021}
\href{https://www.iea.org/reports/heating}{Heating – {Analysis}}, Tech. rep.,
  International Energy Agency.
\newline\urlprefix\url{https://www.iea.org/reports/heating}

\bibitem{MISO2021}
\href{https://cdn.misoenergy.org/Electrification%20Insights538860.pdf}{{MISO}
  {Electrification} {Insights}}, Tech. rep., Midcontinent Independent System
  Operator (Oct. 2021).
\newline\urlprefix\url{https://cdn.misoenergy.org/Electrification%20Insights538860.pdf}

\bibitem{NYISO2020}
\href{https://www.nyiso.com/documents/20142/2226333/2020-Gold-Book-Final-Public.pdf/}{2020
  {Load} \& {Capacity} {Data}}, Tech. rep., The New York Independent System
  Operator, Inc.
\newline\urlprefix\url{https://www.nyiso.com/documents/20142/2226333/2020-Gold-Book-Final-Public.pdf/}

\bibitem{ERCOT2021}
\href{https://www.ercot.com/mktinfo/loadprofile}{Load {Profiling}}, Tech. rep.,
  Electric Reliability Council of Texas.
\newline\urlprefix\url{https://www.ercot.com/mktinfo/loadprofile}

\bibitem{reeve2022}
H.~Reeve, \href{https://www.osti.gov/servlets/purl/1842482/}{The {Distribution}
  {System} {Operator} with {Transactive} ({DSO}+{T}) {Study}}, Tech. Rep.
  PNNL-32170-Sum, 1842482 (Jan. 2022).
\newblock \href {https://doi.org/10.2172/1842482} {\path{doi:10.2172/1842482}}.
\newline\urlprefix\url{https://www.osti.gov/servlets/purl/1842482/}

\bibitem{PUMS2019}
\href{https://www.census.gov/programs-surveys/acs/microdata.html}{Public {Use}
  {Microdata} {Sample}}, Tech. rep., American Community Survey (2019).
\newline\urlprefix\url{https://www.census.gov/programs-surveys/acs/microdata.html}

\bibitem{busby2021}
J.~W. Busby, K.~Baker, M.~D. Bazilian, A.~Q. Gilbert, E.~Grubert, V.~Rai, J.~D.
  Rhodes, S.~Shidore, C.~A. Smith, M.~E. Webber,
  \href{https://linkinghub.elsevier.com/retrieve/pii/S2214629621001997}{Cascading
  risks: {Understanding} the 2021 winter blackout in {Texas}}, Energy Research
  \& Social Science 77 (2021) 102106.
\newblock \href {https://doi.org/10.1016/j.erss.2021.102106}
  {\path{doi:10.1016/j.erss.2021.102106}}.
\newline\urlprefix\url{https://linkinghub.elsevier.com/retrieve/pii/S2214629621001997}

\bibitem{glazer2021}
Y.~R. Glazer, D.~M. Tremaine, J.~L. Banner, M.~Cook, R.~E. Mace,
  J.~Nielsen-Gammon, E.~Grubert, K.~Kramer, A.~M.~K. Stoner, B.~M. Wyatt,
  A.~Mayer, T.~Beach, R.~Correll, M.~E. Webber,
  \href{https://www.worldscientific.com/doi/abs/10.1142/S2345737621500226}{Winter
  {Storm} {Uri}: {A} {Test} of {Texas}’ {Water} {Infrastructure} and {Water}
  {Resource} {Resilience} to {Extreme} {Winter} {Weather} {Events}}, Journal of
  Extreme Events (2021) 2150022\href
  {https://doi.org/10.1142/S2345737621500226}
  {\path{doi:10.1142/S2345737621500226}}.
\newline\urlprefix\url{https://www.worldscientific.com/doi/abs/10.1142/S2345737621500226}

\bibitem{king2021}
C.~W. King, J.~D. Rhodes, J.~Zarnikau, N.~Lin, E.~Kutanoglu, B.~Leibowicz,
  D.~Niyogi, V.~Rai, S.~Santoso, D.~Spence, S.~Tompaidis, H.~Zhu,
  E.~Funkhouser, B.~Austgen, The {Timeline} and {Events} of the {February} 2021
  {Texas} {Electric} {Grid} {Blackouts}  101.

\bibitem{FERC2021}
\href{https://www.ferc.gov/news-events/news/ferc-nerc-staff-review-2021-winter-freeze-recommend-standards-improvements}{{FERC},
  {NERC} {Staff} {Review} 2021 {Winter} {Freeze}, {Recommend} {Standards}
  {Improvements}}, Tech. rep., Federal Energy Regulatory Commission (Sep.
  2021).
\newline\urlprefix\url{https://www.ferc.gov/news-events/news/ferc-nerc-staff-review-2021-winter-freeze-recommend-standards-improvements}

\bibitem{villarreal2021}
M.~Villarreal,
  \href{https://www.cbsnews.com/news/texas-snow-winter-storm-warning-arctic-blast-2021-02-14/}{All
  254 {Texas} counties under winter storm warning as arctic blast heads east}
  (Feb. 2021).
\newline\urlprefix\url{https://www.cbsnews.com/news/texas-snow-winter-storm-warning-arctic-blast-2021-02-14/}

\bibitem{EIA2022}
\href{https://www.eia.gov/state/?sid=TX}{U.{S}. {Energy} {Information}
  {Administration} - {EIA} - {Independent} {Statistics} and {Analysis}}.
\newline\urlprefix\url{https://www.eia.gov/state/?sid=TX}

\bibitem{clack2021}
C.~T.~M. Clack, A.~Choukulkar, B.~Cote, S.~A. McKee, {ERCOT} {Winter} {Storm}
  {Uri} {Blackout} {Analysis} ({February}, 2021), Tech. rep., Vibrant Clean
  Energy (Mar. 2021).

\bibitem{doss-gollin2021}
J.~Doss-Gollin, D.~J. Farnham, U.~Lall, V.~Modi,
  \href{https://iopscience.iop.org/article/10.1088/1748-9326/ac0278}{How
  unprecedented was the {February} 2021 {Texas} cold snap?}, Environmental
  Research Letters 16~(6) (2021) 064056.
\newblock \href {https://doi.org/10.1088/1748-9326/ac0278}
  {\path{doi:10.1088/1748-9326/ac0278}}.
\newline\urlprefix\url{https://iopscience.iop.org/article/10.1088/1748-9326/ac0278}

\bibitem{ERCOT}
\href{https://www.ercot.com/gridinfo/load/load_hist}{Hourly {Load} {Data}
  {Archives}}, Tech. rep., Electric Reliability Council of Texas.
\newline\urlprefix\url{https://www.ercot.com/gridinfo/load/load_hist}

\bibitem{ERCOT2009}
{EILS} in the {CDR}, Tech. rep., Electric Reliability Council of Texas (Dec.
  2009).

\bibitem{FERCandNERC2011}
Report on outages and curtailments during the {Southwest} cold weather event,
  Tech. rep., The Federal Energy Regulatory Commission and the North American
  Electric Reliability Corporation.

\bibitem{ERCOTb}
\href{https://www.ercot.com/mktinfo/data_agg/4cp}{{ERCOT} {Four} {Coincident}
  {Peak} {Calculations}}, Tech. rep., Electric Reliability Council of Texas.
\newline\urlprefix\url{https://www.ercot.com/mktinfo/data_agg/4cp}

\bibitem{EIA2021}
\href{https://www.eia.gov/energyexplained/units-and-calculators/degree-days.php}{Degree-days
  - {U}.{S}. {Energy} {Information} {Administration} ({EIA})}.
\newline\urlprefix\url{https://www.eia.gov/energyexplained/units-and-calculators/degree-days.php}

\bibitem{degreedaysa}
\href{https://www.degreedays.net/calculation}{Calculating {Degree} {Days}}.
\newline\urlprefix\url{https://www.degreedays.net/calculation}

\bibitem{degreedaysb}
\href{https://www.degreedays.net/}{Heating \& {Cooling} {Degree} {Days} –
  {Free} {Worldwide} {Data} {Calculation}}.
\newline\urlprefix\url{https://www.degreedays.net/}

\bibitem{wunderground}
\href{https://www.wunderground.com/history}{Weather {History} \& {Data}
  {Archive} {\textbar} {Weather} {Underground}}.
\newline\urlprefix\url{https://www.wunderground.com/history}

\bibitem{uscensus}
\href{https://www.census.gov/programs-surveys/popest/data/tables.html}{Population
  and {Housing} {Unit} {Estimates} {Tables}}, section: Government.
\newline\urlprefix\url{https://www.census.gov/programs-surveys/popest/data/tables.html}

\bibitem{ERCOTc}
\href{https://www.ercot.com/news/mediakit/maps}{Maps}.
\newline\urlprefix\url{https://www.ercot.com/news/mediakit/maps}

\bibitem{NASAEarthExchange}
\href{https://www.nccs.nasa.gov/services/data-collections/land-based-products/nex-gddp-cmip6}{{NASA}
  {Earth} {Exchange} {Global} {Daily} {Downscaled} {Projections}
  ({NEX}-{GDDP}-{CMIP6}) {\textbar} {NASA} {Center} for {Climate}
  {Simulation}}.
\newline\urlprefix\url{https://www.nccs.nasa.gov/services/data-collections/land-based-products/nex-gddp-cmip6}

\bibitem{abram2019}
Abram, {N}., {J}.-{P}. {Gattuso}, {A}. {Prakash}, {L}. {Cheng}, {M}.{P}.
  {Chidichimo}, {S}. {Crate}, {H}. {Enomoto}, {M}. {Garschagen}, {N}. {Gruber},
  {S}. {Harper}, {E}. {Holland}, {R}.{M}. {Kudela}, {J}. {Rice}, {K}.
  {Steffen}, and {K}. von {Schuckmann}, 2019: {Framing} and {Context} of the
  {Report} {Supplementary} {Material}. {In}: {IPCC} {Special} {Report} on the
  {Ocean} and {Cryosphere} in a {Changing} {Climate} [{H}.-{O}. {Pörtner},
  {D}.{C}. {Roberts}, {V}. {Masson}-{Delmotte}, {P}. {Zhai}, {M}. {Tignor},
  {E}. {Poloczanska}, {K}. {Mintenbeck}, {A}. {Alegría}, {M}. {Nicolai}, {A}.
  {Okem}, {J}. {Petzold}, {B}. {Rama}, {N}.{M}. {Weyer} (eds.)]. {In} press.,
  Tech. rep.

\bibitem{ERCOT2021b}
\href{https://www.ercot.com/files/docs/2021/12/29/CapacityDemandandReservesReport_December2021.pdf}{Report
  on the {Capacity}, {Demand} and {Reserves} ({CDR}) in the {ERCOT} {Region},
  2022-2031}, Tech. rep., Electric Reliability Council of Texas (Dec. 2021).
\newline\urlprefix\url{https://www.ercot.com/files/docs/2021/12/29/CapacityDemandandReservesReport_December2021.pdf}

\bibitem{NRG2021}
\href{https://www.nrg.com/assets/documents/energy-policy/p-52373_nrg-comments-to-september-2-2021-questions.pdf}{{NRG}
  {Energy}, {Inc}.'s {Coments} on the {Commission}'s {September} 2, 2021
  {Questions} for {Comment}}, Tech. rep., NRG Energy.
\newline\urlprefix\url{https://www.nrg.com/assets/documents/energy-policy/p-52373_nrg-comments-to-september-2-2021-questions.pdf}

\bibitem{du2019}
P.~Du, N.~Lu, H.~Zhong, Demand {Responses} in {ERCOT}, in: Demand {Response} in
  {Smart} {Grids}, 2019, pp. 85--119.

\bibitem{EIA2018}
\href{https://www.eia.gov/consumption/residential/data/2015/hc/php/hc6.8.php}{Residential
  {Energy} {Consumption} {Survey} ({RECS})}, Tech. rep., U.S. Energy
  Information Administration.
\newline\urlprefix\url{https://www.eia.gov/consumption/residential/data/2015/hc/php/hc6.8.php}

\bibitem{white2019}
P.~R. White, J.~D. Rhodes, Electrification of {Heating} in the {Texas}
  {Residential} {Sector}, Tech. rep., Ideasmiths, LLC (2019).

\bibitem{white2021}
P.~R. White, J.~D. Rhodes, E.~J. Wilson, M.~E. Webber,
  \href{https://linkinghub.elsevier.com/retrieve/pii/S0306261921005559}{Quantifying
  the impact of residential space heating electrification on the {Texas}
  electric grid}, Applied Energy 298 (2021) 117113.
\newblock \href {https://doi.org/10.1016/j.apenergy.2021.117113}
  {\path{doi:10.1016/j.apenergy.2021.117113}}.
\newline\urlprefix\url{https://linkinghub.elsevier.com/retrieve/pii/S0306261921005559}

\bibitem{2022StatePlan}
\href{https://www.twdb.texas.gov/waterplanning/data/projections/2022/popproj.asp}{2022
  {State} {Plan} {Population} {Projections} {Data} {\textbar} {Texas} {Water}
  {Development} {Board}}.
\newline\urlprefix\url{https://www.twdb.texas.gov/waterplanning/data/projections/2022/popproj.asp}

\bibitem{Lee2022}
J.~Lee,
  \href{https://www.wsj.com/articles/texas-has-an-obvious-affordable-fix-for-its-electricity-problem-11671067175}{Texas
  {Has} an {Obvious}, {Affordable} {Fix} for {Its} {Electricity} {Problem}},
  Wall Street Journal (Dec. 2022).
\newline\urlprefix\url{https://www.wsj.com/articles/texas-has-an-obvious-affordable-fix-for-its-electricity-problem-11671067175}

\bibitem{lewin2022}
D.~Lewin,
  \href{https://www.renewableenergyworld.com/solar/texas-grid-regulator-eyes-big-changes-for-ders/}{Texas
  grid regulator eyes big changes for {DERs}} (May 2022).
\newline\urlprefix\url{https://www.renewableenergyworld.com/solar/texas-grid-regulator-eyes-big-changes-for-ders/}

\bibitem{NERC2022}
2022 {Long}-{Term} {Reliability} {Assessment}, Tech. rep., North American
  Electric Reliabiliaty Corporation (Dec. 2022).

\end{thebibliography}

\end{document}